\documentclass[a4paper,12pt]{article}

\usepackage[centertags]{amsmath}
\usepackage{amssymb}

\usepackage{graphicx}
\usepackage{epsfig}
\usepackage{ulem}
\usepackage[english]{babel}
\usepackage{array}
\usepackage{amsthm}
\usepackage{latexsym}

\usepackage{yfonts}
\usepackage{mathrsfs}
\usepackage[mathcal]{euscript}

\pdfoutput=1
\usepackage{epsfig}
 \usepackage{jheppubm}
\usepackage{mathdots}
\usepackage{MnSymbol}
\usepackage{multirow}

\definecolor{darkraspberry}{rgb}{0.53, 0.15, 0.34}

\definecolor{darkblue}{rgb}{0., 0, 1}

\definecolor{dgreen}{rgb}{0.,0.6,0.}

\newcommand{\be}{\begin{equation}}
\newcommand{\ee}{\end{equation}}
\newcommand{\bea}{\begin{eqnarray}}
\newcommand{\eea}{\end{eqnarray}}

\newcommand{\cI}{{\cal I}}

\title{Schwarzschild black holes, Islands and Virasoro algebra}
\author{Daniil Stepanenko, Igor Volovich}

\affiliation{Steklov Mathematical Institute, Russian Academy of Sciences,\\Gubkina str. 8, 119991, Moscow, Russia}

\emailAdd{dstepanenko@mi-ras.ru, volovich@mi-ras.ru}

\abstract{The Schwarzschild black hole metric with mass $M$ has the limit of the vanishing mass when one get the Minkowski space. We study behavior of the entanglement entropy by using the island formula and the limit of the vanishing black hole mass.
The black hole information problem appears if one considers the lowering its mass. It was noted recently that the formula derived for eternal black holes leads to an increase of the entanglement entropy, which is singular when $M \to 0$. In this process, it arises the other problem of the Schwarzschild black hole explosion, as the black hole temperature blows up as the mass vanishes.We show that it is possible to solve the entropy explosion for small masses, if one admit the following bound: $c<AM$ where $c$ is central charge of the Virasoro algebra, M is black hole mass and A is positive constant. 
We are examining the possibilities of applying such a mechanism to calculate entropy with and without an island. It was shown that applying the mechanism of the dependence of the central charge on the black hole mass allows us to obtain a similar Page curve without using an island.}

\begin{document}
\maketitle

\section{Introduction}
It is well known that the Hawking black hole information problem is that evaporation of black hole contradicts to unitary evolution of quantum mechanics \cite{Hawking:1975vcx,Hawking:1976ra,Frolov:1998wf,Susskind:2005js}.

There is a related problem of explosion of Schawrzschild black hole during evaporation, because that temperature T = 1/8$\pi$M and entropy of radiation blow up for small mass M \cite{Hawking:1974rv,Arefeva:2021byb}.

It is proposed \cite{Arefeva:2022guf}, that for Reissner-Nordstrom and rotating Kerr black holes one can avoid the blow up of temperature and entropy of radiation, if to admit dependence of electro charge and angular momentum on the black hole mass. The Page curve for the entropy of radiation  is obtained, which is usually considered as the step for the solution of the information problem.

An island formula for the entanglement entropy for the black hole is suggested at 
\cite{Penington:2019npb,Almheiri:2019psf,Almheiri:2019hni} aimed to help in understanding the information problem. 

This formula was applied for the eternal Schwarzschild black hole \cite{Hashimoto:2020cas}, but to be appropriate for the information problem, we should use it for the adiabatic evaporation of the Schwarzschild black hole, as it was discussed in \cite{Arefeva:2021kfx}, were there is an increase of the entanglement entropy of  black hole radiation, due to the presence of the term with the central charge in this formula.

The island formula for the Reissner-Nordstrom black hole was considered in \cite{Wang:2021woy,Kim:2021gzd,Arefeva:2022cam}.

For a more detailed discussion of the island formula for the Schwarzschild black hole one might see \cite{Ageev:2022hqc,Ageev:2022qxv}.  

The present paper shows that by allowing the central charge of the Virasoro algebra to depend on the black hole mass, one can avoid the infinite growth of entanglement entropy during evaporation. 

\section{Setup}
The Schwarzschild metric allows the limit of masses to zero, it defines a flat Minkowski spacetime. Consequently, all effects of the Schwarzschild metric should also allow the mass limit to zero. But well, it is known that the Hawking temperature is singular for such small masses, from our point of view the reason is the Kruskal coordinates. Also there remains the question of information loss. Two models can be considered.

\textbf{Model 1}: black hole formed and began to evaporate. It emits a spectrum of particles, as a result of this process we must consider the metric backreaction, and then when it reduces, we must also consider the effects of quantum gravity. But we have no way to describe temporal evolution on the Planck scale.

\textbf{Model 2} : We consider the conformal quantum field in the classical Schwarzschild black hole background. We interpret the classical background as an eternal black hole without taken into account evaporation and backraction. \\The Schwarzschild metric has a fixed mass, but we can choose any mass, including the case of the mass which tends to zero. 
In such a model, we can say that there is a process of evaporation, but the mass of a given black hole is constant, because of the flow of matter from the reservoir. 

And it is this model of black hole evaporation that we will consider. We choose mass as an arbitrary parameter.
This consideration avoids the backreaction problem and Planckian scales. Moreover, in this case we obtain the eternal black holes, which is in agreement with the calculations made in the paper \cite{Hashimoto:2020cas}. 

The Schwarzschild black hole metric we consider as 
\begin{align}
ds^2 = - \,  \frac{r - r_{\rm h}}{r} dt^2 + \frac{r}{r - r_{\rm h}} dr^2 + r^2 d \Omega^2 
\end{align}
with $d\Omega^2=d \theta ^2 +\sin^2\theta  d\varphi^2$ and the Schwarzschild horizon radius $r_{\rm h}=2GM$, where $M$ is mass of the black hole and $G$ is the Newton constant.\\
The Hawking temperature is 
\begin{align}
T_H = \frac{1}{4 \pi r_{\rm h}} = \frac{1}{8\pi GM} 
\end{align}
It has been show in \cite{Arefeva:2021byb}, that the solution of the wave equation for a massless scalar field under Kruskal transformation gives a Planckian distribution with a temperature equivalent to the Hawking temperature.\\
In more detail, you can get  tortoise coordinate $r_*$  by finding a solution to the equation 

  \begin{equation}
    \frac{dr}{dr_*}=(-2M+r) 
\end{equation}
The solution can be written in the form
\begin{equation}
r_*=r+2 M \log \left(\frac{r}{2 M}-1\right)
\end{equation}
Using $r_*$ we can define o Eddincton-Finkelstein coordinates
\begin{equation}
u=t-r_* ~~~~~
v=t+r_*
\end{equation}

(coordinates $u,v$ cover whole $R^2$),
and one has
\begin{equation}
ds_2^{2}=-(1-\frac{2M}{r})dt^2+(1-\frac{2M}{r})^{-1}dr^2=-(1 - \frac{2 M}{r}) du dv\label{ds-uv}
\end{equation}

The Kruskal coordinates are 
\begin{equation}
\label{Kruskal_U_V}
U=-e^{-\frac{u}{4 M}} ~~~~~ V=e^{\frac{v}{4M}}
\end{equation}
and  the Schwarzschild metric becomes 
\begin{equation}
\label{NEW_Schw_K}
ds^2=-\frac{32M^3}{r} \,e^{-r/2M}\,dUdV\,+ r^2d \Omega^2
\end{equation}
where  $r$ is defined from  equation
\begin{equation}
\left(\frac{r}{2 M}-1\right)e^{\frac{r}{ 2M}}=-UV 
\end{equation}

Note that  the Kruskal coordinates \eqref{Kruskal_U_V} and metric \eqref{NEW_Schw_K} are singular in the limit $M\to 0$.

The wave equation  for the massles scalar field $\phi(U,V)$  in these coordinate system are 
\begin{equation}
    \partial_U \partial_V \phi = 0
\end{equation}
They can be represented as combinations of the left and right modes, \\
$\phi(U,V)=\phi_R (U)+\phi_L (V)$.
For the real right mode (for the left mode all consideration is similar and will be omitted) one has
\begin{equation}
\phi _R(U)=\int _0^\infty d\omega (f_\omega a_\omega+f_\omega ^*a^+_\omega),\quad 
f_\omega(U)=\frac{1}{\sqrt{4\pi \omega}}e^{-i\omega a}
\end{equation}
where
\begin{equation}
[a_\omega,a^+_{\omega^\prime}]=\delta(\omega-\omega^\prime)
\end{equation}
The vacuum average in Kruskal coordinates
\begin{equation}
\langle 0_{K}|N_{\omega}(a)|0_{K}\rangle \equiv\langle 0_{K}|a_{\omega}^+a_{\omega}|0_{K}\rangle
=\int_0^{\infty} d\mu \,|\beta_{\omega\mu}|^2
\end{equation}
Calculation of the Bogoliubov coefficient in Kruskal coordinates $\beta_{\omega\nu}$ leads to the Planck distribution
\begin{equation}
|\beta_{\omega\,\mu}|^2=\frac{2M}{\pi \mu}\frac{1}{e^{8\pi M \omega }-1}
\end{equation}
with the temperature
\begin{equation}
    T=\frac{1}{8\pi M}
\end{equation}

From these simple calculations it is clearly seen that the black holes radiated like black body with temperature $T_H$ and  explosion of temperature in the Schwarzschild metric is observed because of the coordinates of the kruskal. In spite of this divergence, we will be interested in the question how entropy behaves if we choose the value of the orbital parameter M to be zero.

 The entropy of radiation $S_R \sim T^3 $ \cite{Frolov:1998wf,Susskind:2005js} that grows for small black hole mass. Here we meet a similar problem, observing an explosion of temperature, we inevitably get an explosion for the entropy of radiation.
The same problem is found when the entanglement entropy of radiation is derived by using the island formula.

In paper \cite{Hashimoto:2020cas}, an expression for the entanglement entropy in 
Schwarzschild black
holes in four spacetime dimension with only free massless matter fields was derived, given by the formula:
\begin{equation}
\label{23}
S_{\cI} = \frac{2\pi r_h^2}{G} + \frac{c}{6} \frac{b-r_h}{r_h}
 + \frac{c}{6}\log \frac{16r_h^3(b-r_h)^2}{G^2b}
\end{equation}
Here $b$ is the boundaries of the radiation region and $c$ is the central charge of the matter fields.

In this paper we take the view point that the formula \eqref{23} can be applied not only to an eternal black hole, but also to an adiabatically evaporating Schwarzschild black hole.

It \cite{Arefeva:2021kfx} has been demonstrated that the island configuration cannot provide a zero entropy value at the end of Schwarzschild black hole evaporation.
The formula \eqref{23} leads to the explosive entropy growth due to the term  $c b/r_h=c b/2MG$, which becomes dominant for small M.

Therefore, one would think, that the island formula does not help in the resolution of the information problem, however in this paper we suggested a way to improve this situation by making the central charge dependent on the black hole mass.

In the case when black hole evaporates adiabatically, due to the finite matter contribution, radiation entropy does not go to zero when the black hole radius goes to zero. 
\\
The
mutual information between two separated region A and B, connected with island I(A; B) is defined by two-dimensional massless
fields,
\begin{equation}
    I(A;B)=-\frac{c}{3}\log{d(x,y)}
\end{equation}
where c is the central charge and $d(x, y)$ is the distance between $x$ and $y$ which are
the boundaries of A and B.
\\
The Virasoro algebra is defined by the commutation relations\cite{Tong:2009np,Bowcock:1987mw,Goddard:1986bp,Chodos:1973gt,Kac:2013bl} : 
\begin{equation}
[L_m,L_n]=(m-n)L_{m+n} + \frac{c}{12}(m^3-m) \delta_{m,-n} 
\end{equation}
where $m$ and $n$ range over integers and the central charge c
commutes with all the other generators $L_n$.
\\
In this note we show that if the central charge satisfies the boundaries, then a finite value of entropy can be obtained.
\begin{equation}
c< A M
\end{equation}

\section{Central charge dependence on the mass of the black hole}
We assume the complete evaporation of black holes. Therefore, to solve the information paradox, we need to reproduce the Page curve. For this purpose we make the assumption that the central charge in our theory depends on the change of mass. We consider this action to be valid and see that the formula \eqref{23} is also applicable to the evaporated Schwarzschild black hole. We assume that the black hole mass changes adiabatically. This process is very slow, comparable with the lifetime of the black hole itself.

We consider a special case with relationship between the central charge and mass in the form 
\begin{equation}
\label{26}
c(M)=A M^{1+\epsilon}
\end{equation}
Where A is a constant and $\epsilon$ is a positive number parameter, $ \epsilon >0$. \\ 
In that case, we have:
\begin{equation}
\label{32}
S_{\cI} = \frac{2\pi (2GM)^2}{G} + \frac{AM^{1+\epsilon}}{6} \frac{b-2GM}{2GM}
 + \frac{AM^{1+\epsilon}}{6}\log \frac{16(2GM)^3(b-2GM)^2}{G^2b}
\end{equation}
If one assumes that the mass monotonically decreases with time, then at the late time (complete evaporation) as M → 0. \\ 
Calculate $ \lim_{M \to 0} S_{\cI}  $ and for make our point clear, take the $\epsilon=1$, now we have 
\begin{equation}
\lim_{M \to 0} S_{\cI} =0
\end{equation}
Thus, we obtain a zero value of entropy at the moment of complete evaporation, which fully coincides with the behavior of the Page curve.
Now we will be interested in the question how the central charge appears in our theory. On what it depends and what properties it has.

\section{Virasoro algebra with $c<1$}

The Virasoro algebra is defined by the commutation relations  \cite{Bowcock:1987mw,Goddard:1986bp,Kac:2013bl}: 
\begin{equation}
[L_m,L_n]=(m-n)L_{m+n} + \frac{c}{12}(m^3-m) \delta_{m,-n} 
\end{equation}
\begin{equation}
[c,L_n]=0
\end{equation}
where $m$ and $n$ range over integers and the central charge $c$ commutes with all the other generators $L_n$, and therefore it can be assumed to take a numerical value in any irreducible
representation.
\subsection{Oscillator Representation of Virasoro Algebra}
Define operator harmonic representations of the Virasoro algebra in Fock space in form \cite{Chodos:1973gt,Chodos:1974je,Kato:1985vq} for n=1,2,3... and $\omega$ the real number:
\begin{equation}
L_n(\omega)=(a_0 + \omega n )a_n 
+ \sum_{j=1} ^{\infty} a^+ _j a_{n+j} 
+ \frac{1}{2}\sum_{j=1} ^{n-1} a_j a_{n-j}
\end{equation}
\begin{equation}
L_{-n}(\omega) = (a_0 - \omega n) a^+_n 
+ \sum_{j=1} ^{\infty} a^+ _{n+j} a_{j} 
+ \frac{1}{2}\sum_{j=1} ^{n-1} a^+_j a_{-n+j}
\end{equation}

\begin{equation}
L_{0}(\omega) = \frac{1}{2}(a_0^2 - \omega^2) 
+ \sum_{j=1} ^{\infty} a^+ _{j} a_{j} 
\end{equation}
Annihilation and creation operators satisfy commutation relations, for $ k=0, 1, 2, ...$ : 
\begin{equation}
 [a_k,a^+_m]=k \delta _{k,m}
\end{equation}
We consider the theory with the vacuum vector $\varphi_0$ : 

\begin{equation}
a_k |\varphi_0\rangle=0 ~~~\text{and}~~~
\langle \varphi_0 | a^+_k=0
\end{equation}
That leads us to the result that the central charge is connected with $\omega$ number parameter in form \cite{Kato:1986rq} :
\begin{equation}
[L_m,L_n]=(m-n)L_{m+n} + \frac{c}{12}(m^3-m) \delta_{m,-n} 
\end{equation}
\begin{equation}
    c=1-12\omega^2
\end{equation}
Real value $\omega$ describe the region c $ \leq $ 1.
This means that Virasoro's algebra allows the dependence of the central charge on the black hole mass \eqref{23}. Now we will be interested in the possibilities of application of this mechanism.

\section{Entropy with and without island }
\subsection{Entropy corresponding to non-island configuration}
Entanglement entropy at late times, in the case of the absence of the island, been derived in \cite{Hashimoto:2020cas}.
\begin{equation}
S_{N\cI}=\frac{c}{6}\frac{t}{r_h}
\label{51}
\end{equation}
At later times the expression \eqref{51} goes to infinity. This is a clear contradiction with the finite value of von Neumann entropy for a finite-dimensional black hole system. To solve this problem, let us apply the mechanism of dependence of the central charge on the black hole mass. \footnote{ We are grateful to Timofei Rusalev for his suggestion to apply the mechanism of dependence of the central charge on mass, to the calculation of entropy without an island.}
\begin{equation}
{S_{N\cI}
 =\frac{AM^{1+\epsilon}}{6 } \frac{t}{r_h} 
 =\frac{\left(\frac{M}{M_0}\right)^{1+\epsilon} t }{12G M}} =\frac{M^{\epsilon} t }{12 M_0^{1+\epsilon} G}
\end{equation}
We choose constant $A=(\frac{1}{M_0})^{1+\epsilon}$, where $M_0$ is initial black mass, so that the value of entropy was a dimensionless quantity.

Now apply the time evolution formula for mass from  \cite{Page:1976df}.
\begin{equation}
    \frac{dM}{dt}= -\frac{1}{G^2} \frac{\alpha}{M^2}
\end{equation}
The solution of this equation is obtained as 

\begin{equation}
\label{54}
    M(t) =\left(\frac{3(G^2 B-\alpha t)}{G^2} \right)^{1/3} = \left(\frac{3(G^2 M_0^3-\alpha t)}{G^2} \right)^{1/3} 
\end{equation}

Where constant of integration was chosen as follows $B = M_0 ^3$. This gives the final expression for entropy without an island.

\begin{equation}
    S_{N\cI}= \frac{ \left(\frac{3(G^2 M_0^3-\alpha t)}{G^2}\right)^{\boldsymbol{\epsilon}/3} t }
    { 12 M_0^{1+\epsilon} G } 
    = \frac{\left(3(M_0^3-\frac{\alpha t}{G^2})\right)^{\boldsymbol{\epsilon}/3}t}
    {12 M_0^{1+\epsilon} G^{\frac{1+2\epsilon}{3}}}
\end{equation}

\begin{figure}
\includegraphics[scale=0.83]{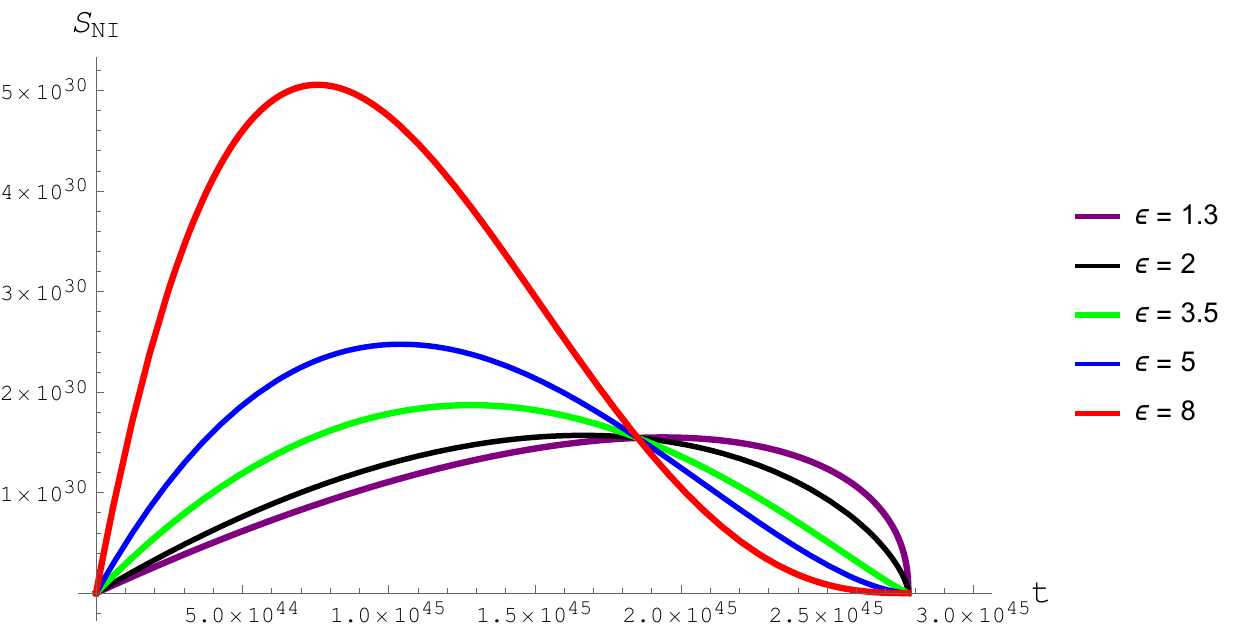}   \\
      \textcolor{blue}{Figure A.} 
  {We observe an increase in entropy without an island, but then a decrease to zero at the end of evaporation.  }
  \label{Fig.A}
\end{figure}

The behavior of entropy under a non-island configuration is shown in \textcolor{blue}{Fig.A}(\ref{Fig.A}).
We can clearly see that the entropy without an island becomes finite if we allow the dependence of the central charge on the mass of the black hole.

Where the initial black hole mass was chosen as $M_0=1\times10^{14}$, $G=1$ and the parameter $\alpha = 3.6 \times 10^{-4}$.

\newpage
\subsection{Entropy corresponding to island configuration}
\begin{figure}
\includegraphics[scale=0.83 ]{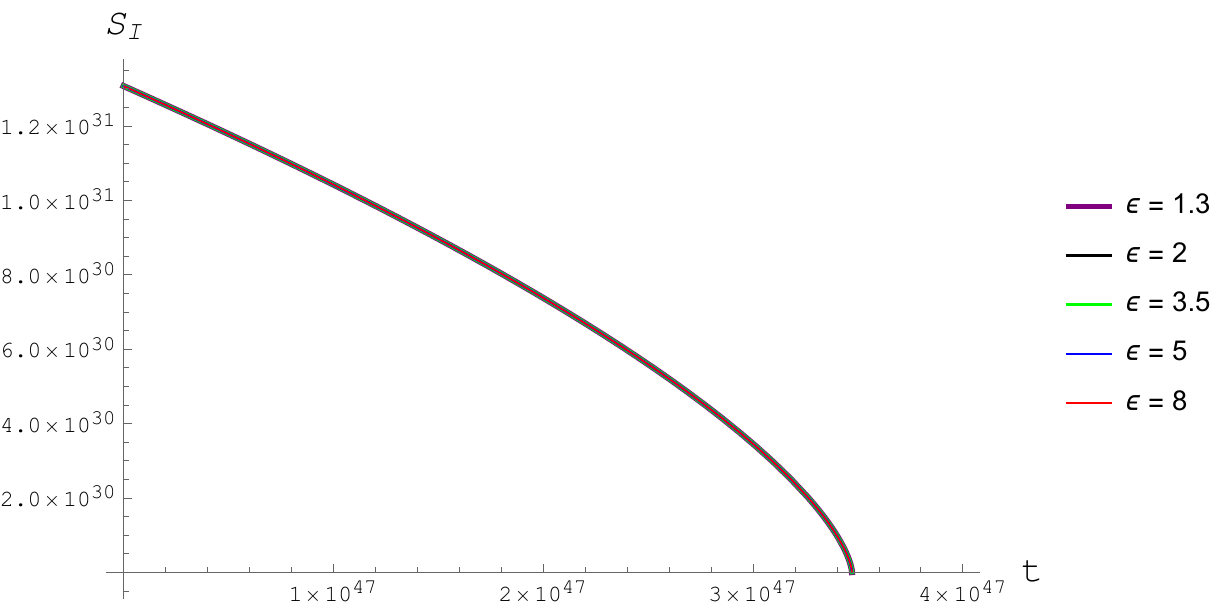}   \\
       \textcolor{blue}{Figure B.} 
 {The entropy with island decreases in the end of evaporation and
 
 becomes finite. And this expression do not depend of the choice $\epsilon$.  }
  \label{Fig.B}
\end{figure}

Let's look at the behavior of entropy under the island configuration \eqref{32}  and apply the mechanism of dependence of the central charge on the black hole mass  \eqref{23}.
\begin{equation}
    S_{\cI} = \frac{2\pi (2GM)^2}{G} 
+ 
\frac{M^{\epsilon}}{6 M_0^{1+\epsilon}} \frac{b-2GM}{2G}
+ 
 \frac{M^{1+\epsilon}}{6 M_0^{1+\epsilon}}\log \frac{16(2GM)^3(b-2GM)^2}{G^2b}
\end{equation}
An expression describing the evolution of mass over time \eqref{54}.
\begin{equation}
    M(t) = \left(3(M_0^3 - \frac{\alpha t}{G^2})\right)^{1/3}
\end{equation}
The final expression for entropy with an island can be written in the form \begin{equation}
    \begin{aligned}
        S_{\cI} = \frac{\left(3 \left(M_0 ^3-\frac{\alpha  t}{G^2}\right)\right)^{\frac{\epsilon +1}{3}}}{6 M_0 ^{\epsilon +1}} \log \frac{128 G \left(3 \left(M_0 ^3-\frac{\alpha  t}  {G^2}\right)\right) \left(b-2 G \sqrt[3]{3 \left(M_0 ^3-\frac{\alpha  t}{G^2}\right)}\right)^2}{b}+ \\
        \frac{(b-2 G \sqrt[3]{3 \left(M_0 ^3-\frac{\alpha  t}{G^2}\right)}) \left(3 \left(M_0 ^3-\frac{\alpha  t}{G^2}\right)\right)^{\epsilon /3}}{12 G M_0 ^{\epsilon +1}} + 8 \pi  G (3 \left(M_0 ^3-\frac{\alpha  t}{G^2}\right))^{2/3}
        \end{aligned}
\end{equation}
The behavior of entropy with an island configuration is shown in \textcolor{blue}{Fig.B}(\ref{Fig.B}).
The graph shows that the island formula using the mechanism of dependence of the central charge on the black hole mass provides a decrease of entropy to zero value. 

Where the initial black hole mass was chosen as $M_0=1\times10^{14}$, $G=1$,the parameter $\alpha = 3.6 \times 10^{-4}$ and island boundaries $b=5$.

\newpage
\section{Discussion and conclusion}
We consider a special constrain on central charge dependence of black hole mass, for the possibility to resolve the singularity of the entanglement entropy using Island configurations for Schwarzschild black holes. The representation of the Virasoro algebra for central charge in region $c < 1$ have been discussed.Central charge goes to zero for small black hole mass. 

This approach solves the problem of the entropy explosion raised in \cite{Arefeva:2021kfx}. We present a mechanism based on the identification of central charge for algebra Virasoro and black hole mass. This makes it possible to, avoiding the singular term for expressing entanglement entropy with island configuration \eqref{23}. We want to make a remark on vanishing matter field. From our understanding, if we consider a theory that contains a Schwarzschild black hole and its mass decreases, then its effect on space time decreases as well. We expect to see an evolution from the Schwarzschild metric to the Minkowski metric. 

As a result of increases of temperature the external observer will see more massive quanta emitted by black hole. In other word, we conclude island formula under constrain \eqref{26} do solve the blow up of the entropy for the Schwarzschild black holes. 

And not only that, application of such mechanism for the none island formula, so provides a finite value of entropy at the end of black hole lifetime.

\section*{Acknowledgement}
We would like to thank Irina Aref'eva, Elizaveta Kuprina and Timofei Rusalev for very helpful discussions.

This work is supported by the Russian Science Foundation(19-11-00320, Steklov Mathematical Institute).

\end{document}